\documentclass[a4paper,12pt]{article}
 \usepackage[left=2cm,right=2cm]{geometry}
\usepackage[utf8x]{inputenc}
\usepackage{amsfonts,amsmath,amssymb,bm,eucal}
\usepackage{graphicx}
\usepackage{epstopdf}

\title{\vspace*{-2cm}
\begin{flushright}
\normalsize{BARI-TH/2014-691}
\end{flushright}
\vspace*{1.5cm}Gluon Condensate at finite temperature and density in holographic QCD}
\author{F.~Giannuzzi\thanks{floriana.giannuzzi@ba.infn.it}\\~\\ 
\normalsize\emph{Dipartimento di Fisica, Universit\`a degli Studi di Bari, via Amendola 173, 70126 Bari, Italy}
}
\date{}

\begin{document}

\maketitle

\begin{abstract}
The lowest dimensional gluon condensate $G_2$ is analysed at finite temperature and chemical potential using a bottom/up holographic model of QCD. Starting from the free energy of the model, pressure, entropy and quark density are obtained. Moreover, at zero chemical potential, the temporal and spatial Wilson loops at low temperature are computed; they are related to the (chromo-)electric and magnetic components of $G_2$, respectively.

\vspace*{0.3cm}
\noindent{pacs:
11.25.Tq, 
11.10.Kk, 
11.15.Tk 
12.38.Lg 
}
\end{abstract}
\section{Introduction}
\label{intro}
The gluon condensate $G_2$ is the vacuum expectation value of the operator $\alpha_s/\pi G^a_{\mu\nu} G^{a,\mu\nu}$, where $G^a_{\mu\nu}$ is the gluon field strength tensor.
Some estimates have been obtained so far for this nonperturbative property of Quantum Chromodynamics (QCD), leading to the value $G_2\simeq0.012$ GeV$^4$, which is however affected by large uncertainties \cite{gcond}.
The behaviour of gluon condensate at finite temperature and density can also be scrutinised to obtain information about QCD in these conditions.
Here, this task will be accomplished through a bottom/up approach to the AdS/QCD correspondence.\\
The gauge/gravity duality has opened a new way of studying QCD, in which nonperturbative calculations are performed within a semiclassical perturbative theory in a 5-dimensional  (5$d$) curved spacetime.
To this aim, in the last decade, some phenomenological models have appeared, in which an effective Lagrangian is constructed in a 5$d$ Anti-de Sitter space (AdS) by following two main guidelines: taking into account the dictionary of the AdS/CFT correspondence \cite{Witten:1998qj}, and trying to mimic well-known QCD properties.
Such a dictionary establishes how to relate quantities of QCD with the ones of the 5$d$ theory.
It claims that a 4$d$ local gauge-invariant operator is dual to a specific field of the supergravity theory. 
Moreover, it also states that the generating functional of the gauge theory ($Z_{CFT}$) is equal to the partition function of the gravity theory ($Z_{S}$), which, in the supergravity limit takes a very simple form:
\begin{equation}\label{eq:adscft}
 Z_{CFT}=Z_S=e^{iS} \,,
\end{equation}
where $S$ is the on-shell action of the supergravity theory.
This allows us to straightforwardly compute QCD correlation functions, {\it i.e.} by deriving a classical action \cite{Witten:1998qj}.

\section{The model}
\label{sec-1}
In the AdS/QCD correspondence, an effective 5$d$ Lagrangian has to include a mass scale to break conformal symmetry. 
In the soft-wall model \cite{Karch:2006pv} this is done by inserting a factor $e^{-\phi(z)}$ in the action, with $\phi(z)=a^2z^2$. With this choice, Regge trajectories for mass spectra are generated, and many analytic calculations at zero temperature and density can be performed (see, {\it e.g.}, \cite{Karch:2006pv,sw,Andreev:2007vn}).

\begin{figure}
 \begin{minipage}[b]{7.5cm}
 \centering
\includegraphics[width=7cm]{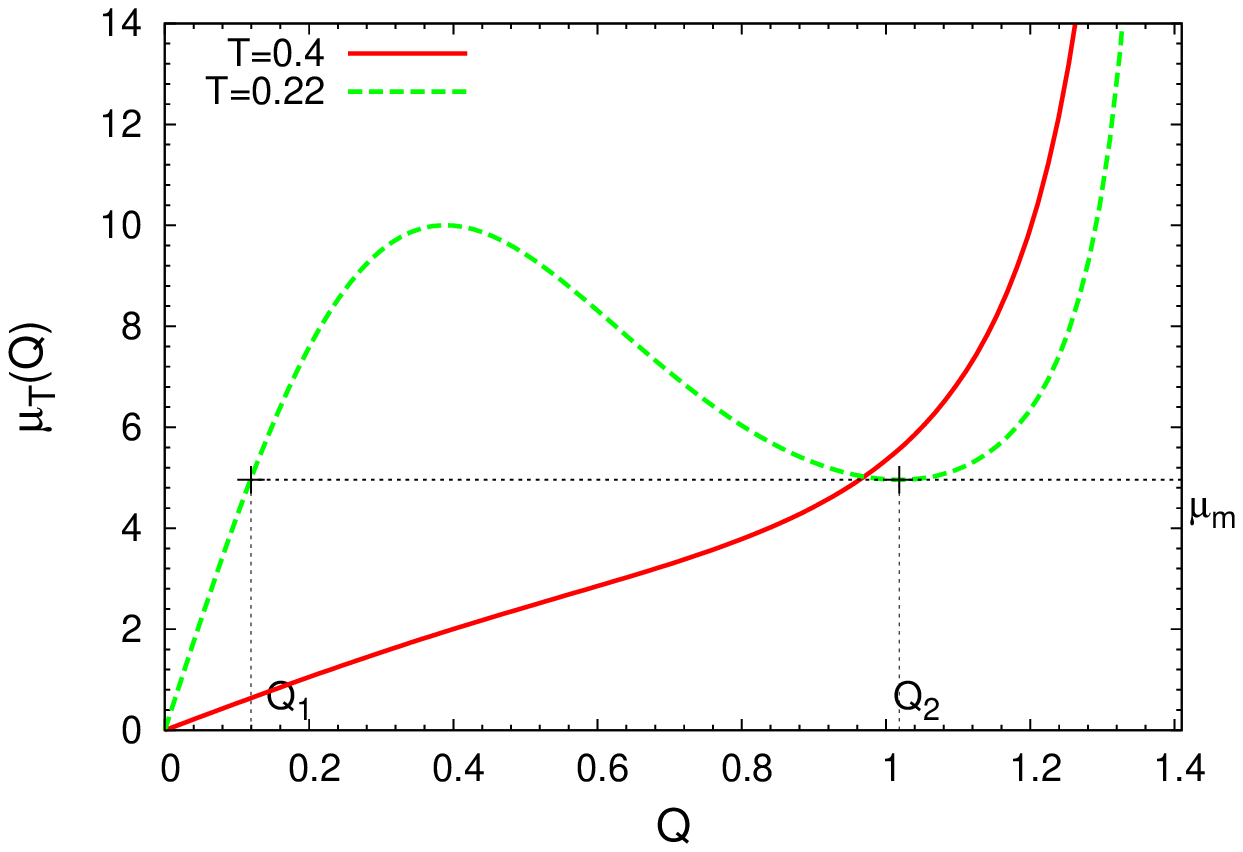}
 \caption{Chemical potential $\mu_T$ at fixed temperature versus $Q$. $c=1$, $g_5^2=1$, $a_E\simeq -2.5$ have been set. }
 \label{fig:muT}  
 \end{minipage}
 \ \hspace{3mm} \hspace{5mm} \
 \begin{minipage}[b]{7.5cm}
 \centering
\includegraphics[width=7cm]{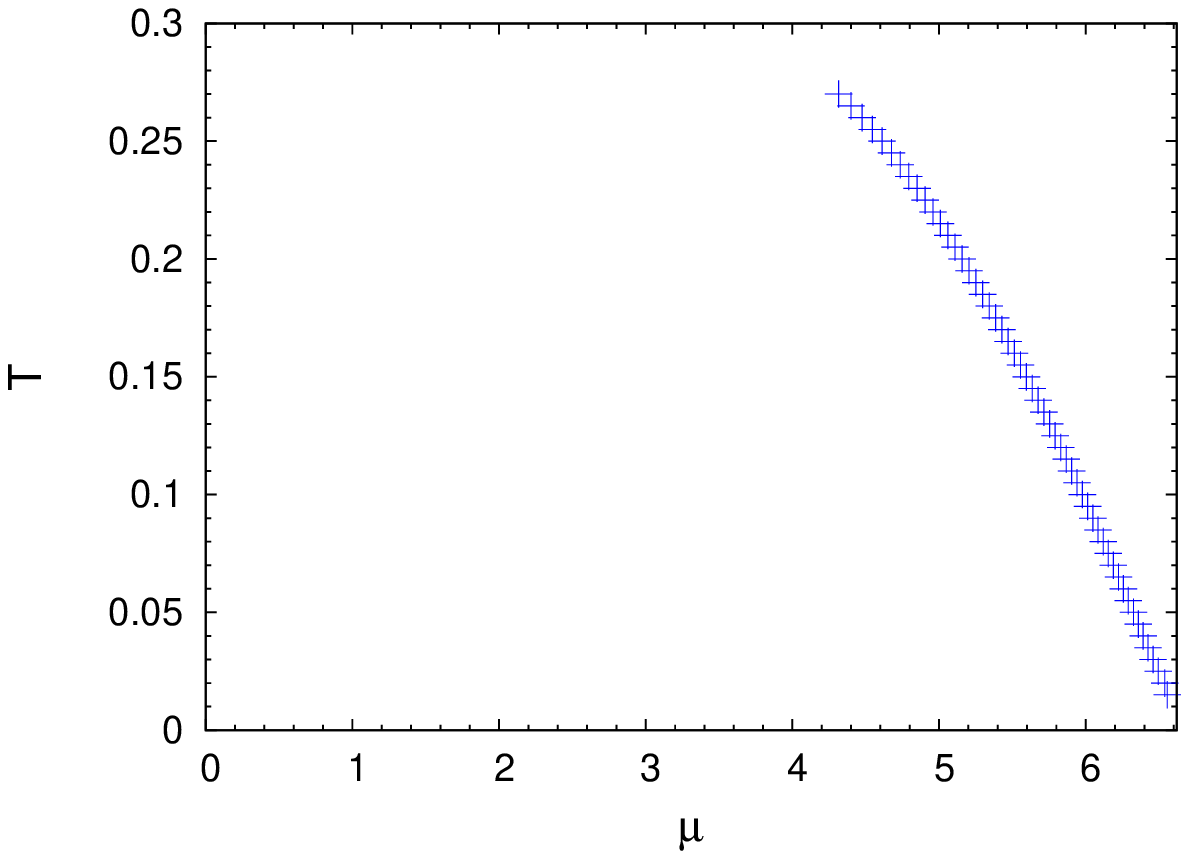}
 \caption{$(T, \mu)$ pairs corresponding to a jump in Fig. 1, and where the thermodynamical quantities present a discontinuity.}
 \label{fig:PD}    
 \end{minipage}
 \end{figure}

Temperature and density are introduced by adding a charged black hole to the AdS space.
The resulting metric in Euclidean space is:
\begin{equation}\label{eq:LineEl}
 ds^2=\frac{R^2}{z^2} \left(f(z)d\tau^2+d\bar x^2+\frac{dz^2}{f(z)} \right) \,\,\, ,
\end{equation}
with
$ f(z)=1-\left( \frac{1}{z_h^4}+q^2 z_h^2 \right) z^4+q^2 z^6$ .
$R$ is the radius of the AdS space ($R=1$ will be considered), $q$ the black-hole charge, $z_h$ the position of the outer horizon of the black hole, defined by the condition $f(z_h)=0$. As a consequence, the holographic coordinate $z$ lies in the interval $0<z<z_h$, being $z=0$ the boundary of the AdS space.
The quark-number operator $q^\dagger q$ is dual to the temporal component of a U(1) gauge field $A_0(z)$, whose boundary value $A_0(0)$ must be equal, according to the dictionary of the correspondence \cite{Witten:1998qj}, to the source of the operator, namely the chemical potential $\mu$.\\
Let us consider the soft-wall model, with dilaton $\phi(z)=-a_E^2 c^2 z^2$, where $c$ represents the mass scale and $a_E$ is an additional parameter. The expression for the gauge field can be found by solving the corresponding equation of motion, as in \cite{Colangelo:2013ila}, getting
\begin{equation}\label{eq:solA0}
 A_0(z)=i \left(\mu-\frac{ \sqrt{3 g_5^2}\, q}{a_E\, c^2} \left(1-e^{-a_E c^2 z^2}\right)\right) \,,
\end{equation}
where $g_5^2$ is a coefficient appearing in the Lagrangian \cite{Karch:2006pv}.
Since $A_0(z)$ must vanish on the black-hole horizon ($A_0(z_h)=0$), the following constraint turns out:
\begin{equation}\label{eq:mu}
 \mu=\frac{\sqrt{3 g_5^2} \, q}{ a_E\, c^2} \left( 1-e^{-a_E c^2 z_h^2} \right) =\frac{\sqrt{3 g_5^2} \, Q}{a_E \, c^2 \, z_h^3} \left( 1-e^{-a_E c^2 z_h^2} \right) \,.
\end{equation}
Temperature is defined by
\begin{equation}\label{eq:temperature}
 T=\frac{1}{4\pi}\left| \frac{df}{dz} \right|_{z=z_h}=\frac{1}{\pi z_h} \left( 1-\frac{q^2 z_h^6}{2} \right)=\frac{1}{\pi z_h} \left( 1-\frac{Q^2}{2} \right)\,,
\end{equation}
where $Q=q z_h^3$, $0\leqslant Q\leqslant \sqrt{2}$.
Eqs. \eqref{eq:mu}-\eqref{eq:temperature} relate temperature and chemical potential to the two black-hole parameters $z_h,q$.
An interesting property of the model can be obtained from these relations.
If one computes $z_h$ as a function of $T$ from Eq.~\eqref{eq:temperature} and uses this expression in Eq.~\eqref{eq:mu}, the chemical potential can be studied at fixed temperature at varying the charge $Q$ in the interval   $0\leqslant Q\leqslant \sqrt{2}$. The result is shown in Fig.~\ref{fig:muT}. As long as $T$ is high, $\mu$ is a monotonic function of $Q$, so a particular value of $\mu$ is reproduced by one particular value of $Q$; this is the case of the $T=0.4 c$ curve in the figure. At lower $T$ (as for $T=0.22 c$ in the figure), the function is not monotonic, and, in some cases, more values of $Q$ give the same chemical potential, but only one has to be chosen. 
We choose the value giving the lowest free energy, corresponding, in Fig.~\ref{fig:muT}, to the branches $0\leqslant Q\leqslant Q_1$ and $Q_2\leqslant Q\leqslant \sqrt{2}$. Therefore, for some low values of temperature, there is a jump from $Q_1$ to $Q_2$ in thermodynamic functions, representing a first order phase transition. The values of $T$ and $\mu$ at which such a jump occurs are collected in Fig.~\ref{fig:PD}.

\section{The gluon condensate}
\label{sec-2}
We have considered two possible computations of the gluon condensate in medium.\\
In the first case, which will be discussed in next subsection, we have computed it by exploiting the trace anomaly equation, which, for massless quarks, reads
\begin{equation}
\Theta^\mu_\mu= \frac{\beta(\lambda)}{\lambda} G^a_{\mu \nu} G^{a, \mu \nu}\,, \label{tranomaly}
\end{equation}
$\Theta^\mu_\mu$ being the trace of the QCD energy momentum tensor, $\displaystyle \lambda=N_c\frac{ g_s^2}{4 \pi}$ the 't Hooft coupling ($N_c$ is the number of colours), and $\beta(\lambda)$ the $\beta$-function of QCD. Standing Eq.~\eqref{tranomaly}, the gluon condensate is given by the vacuum value of the trace of the QCD energy momentum tensor (right hand side), hence it can be obtained from thermodynamic functions.\\
In the other case, studying a small circular Wilson loop will give information about the chromo-electric and magnetic components of the gluon condensate at small temperatures.

\subsection{Gluon condensate from trace anomaly equation}
\begin{figure}
  \begin{minipage}[b]{7.7cm}
 \centering
\includegraphics[width=7cm,clip]{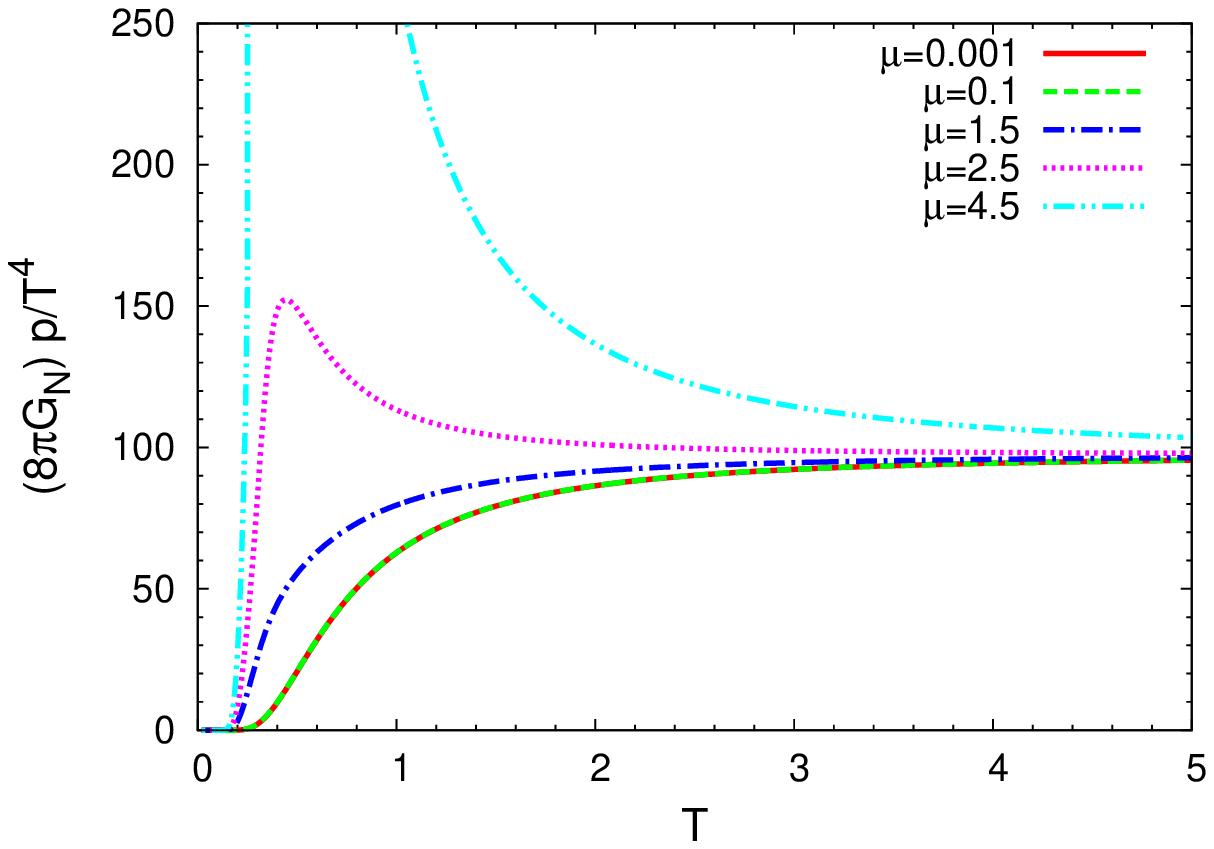}
 \caption{$p(T,\mu)/T^4$ versus $T$, for some values of chemical potential ($c$ units); $g_5^2=1$.}
 \label{fig:press}    
\end{minipage}
 \ \hspace{3mm} \hspace{3mm} \
 \begin{minipage}[b]{7.5cm}
 \centering
\includegraphics[width=7cm,clip]{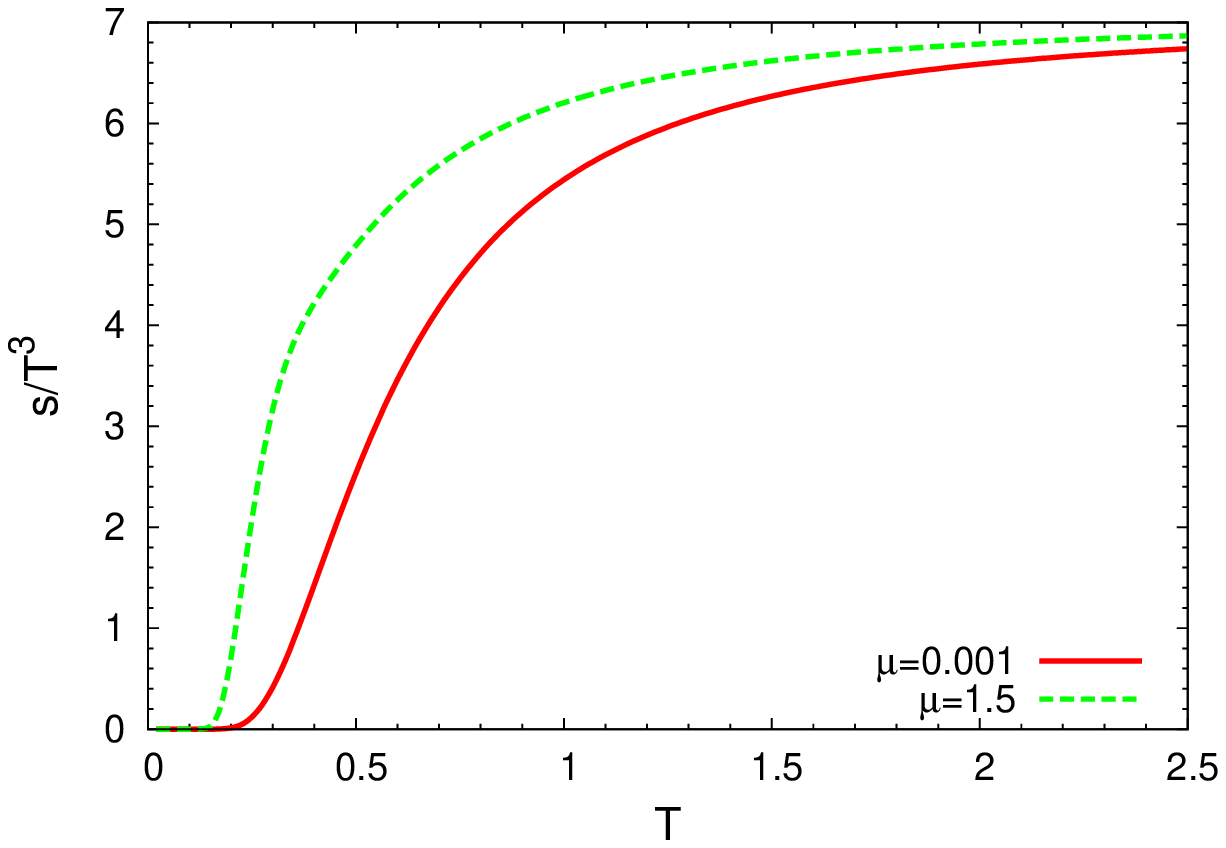}
 \caption{Entropy density, divided by $T^3$, versus temperature $T$ ($c=1,g_5^2=1$).}
 \label{fig:entr} 
  \end{minipage}
 \end{figure}

\begin{figure}
 \centering
\includegraphics[width=7.5cm,clip]{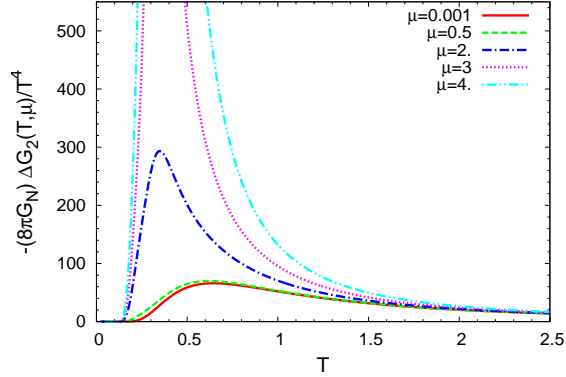}
 \caption{Gluon condensate, divided by $T^4$, versus temperature $T$ ($c=1,g_5^2=1$).}
 \label{fig:g2}       
 \end{figure}

Eq.~\eqref{eq:adscft} establishes a relation between the $4d$  partition function $Z_{CFT}$ and the $5d$ gravity action, allowing us to compute the free-energy density from
\begin{equation}\label{eq:adscft2}
{\cal F}=-\frac{T}{V}\log {Z_{CFT}}=\frac{T}{V} S_E\,,
\end{equation}
where $S_E$ is the on-shell action in the Euclidean space.
Then, the free energy is given by  \cite{Colangelo:2013ila}:
\begin{eqnarray}\label{eq:freeen}
 {\cal F} &=& -\frac{1}{16 \pi G_N}\int_0^{z_h} dz\, \sqrt{g}\, e^{a_E c^2 z^2} \left( {\cal R}-2\Lambda-\frac{1}{4g_5^2} F^2\right) \nonumber \\
 &=& -\frac{1}{16 \pi G_N}\int_0^{z_h} dz\, \frac{e^{a_E c^2 z^2}}{z^5} \left( -8 -2 Q^2 \frac{z^6}{z_h^6} -\frac{1}{2g_5^2} z^4 A_0'(z)^2\right) \,;
\end{eqnarray}
$G_N$ is the Newton constant in $5d$. It has two contributions, coming from the Einstein-Hilbert and  the Maxwell terms of the action.
Eq.~\eqref{eq:freeen} needs to be regularised: by subtracting the divergent terms $1/\epsilon^4$, $2 a_E c^2/\epsilon^2$ and $- a_E^2 c^4 \log(c^2\epsilon^2)$, the final expression is \cite{Colangelo:2013ila}
\begin{eqnarray}\label{eq:subtr}
{\cal F}(z_h,Q) &=& \frac{1}{8\pi G_N} \left( -\frac{e^{ a_E c^2 z_h^2}}{z_h^4} -\frac{1}{2} a_E^2 c^4\left( -3 +2 \gamma_E-2 \Gamma(-1,- a_E c^2 z_h^2) + \log( a_E^2) \right)\right. \nonumber\\
&& \left. +\frac{Q^2}{2 a_E\, c^2\, z_h^6} \left(e^{a_E c^2 z_h^2}-1\right)-\frac{3 Q^2}{2 a_E \, c^2\, z_h^6} \left( 1-e^{-a_E\, c^2\, z_h^2}\right)\right)\,.
\end{eqnarray}
Once the free energy is fixed, all the thermodynamic functions can be computed:
\begin{itemize}
 \item pressure: $p=-{\cal F}$;
 \item entropy density: $s=\partial p/\partial T$;
 \item quark density: $\rho=\partial p/\partial \mu$;
 \item  energy density: $\epsilon=Ts-p+\mu\rho$.
\end{itemize}
Then, by means of the trace anomaly equation, the gluon condensate is obtained \cite{Leutwyler1992cd}:
\begin{equation}
 \Delta G_2(T,\mu)=G_2(T,\mu)-G_2(0,0)= -\epsilon(T,\mu)+ 3 p(T,\mu)\,.
\end{equation}
The numerical results for pressure, entropy density and gluon condensate are shown in Figs.~\ref{fig:press},\ref{fig:entr},\ref{fig:g2}.
In the limit $(\mu ,T)\to 0$, the pressure behaves as
\begin{equation}
p(T,\mu) \to \frac{1}{8\pi G_N} \frac{1}{2} a_E^2\, c^4\, (2 \gamma_E-3+2 \log(- a_E)) \,\,\,; \label{press1}
\end{equation}
this fixes the parameter $a_E$, since a vanishing pressure can be obtained only if $a_E=-e^{3/2-\gamma_E}\sim -2.5$.
The mass scale of the model $c$ is fixed by the $\rho$ meson mass, as done in \cite{Karch:2006pv}, finding $c\sim 0.25$ GeV.
For small values of the chemical potential, $p/T^4$ has a monotonic $T$ dependence; all the functions are independent of $\mu$ at high values of $T$.
It is worth emphasising that in the present model the entropy density vanishes at vanishing temperature even when the chemical potential is finite  (for $\mu\lesssim 6.8$), a property which is not shared by other modified versions of the Reissner-Nordstr\"om model \cite{Ammon2011hz} (unless a Hawking-Page transition is assumed).

\subsection{Gluon condensate from Wilson loop}\label{sec:wloop}
Another way of computing $G_2$ is through a small Euclidean Wilson loop. 
The vacuum expectation value of this object ($\langle W({\cal C})\rangle$) can be expanded in powers of the area $s$ of the loop \cite{Di Giacomo1981wt},
\begin{equation}\label{eq:logW}
\log \left( \langle W \rangle \right) =-\sum_n c_n \alpha_s^n-\frac{\pi^2}{36}Z G_2 s^2 +{\cal O}(s^3)\,.
\end{equation}
The first term is a perturbative series in $\alpha_s$, and  $Z$ is a renormalisation constant; the gluon condensate $G_2$ appears in the ${\cal O}(s^2)$ term of the expansion.
Eq.~\eqref{eq:logW} can be studied in the holographic approach by using the following identification \cite{Maldacena:1998im}
\begin{equation}\label{eq:expNG}
\langle W({\cal C}) \rangle \sim e^{-S_{NG}}\,,
\end{equation}
where $S_{NG}$ is the Nambu-Goto action:
\begin{equation}\label{eq:NGaction}
S_{NG}=\frac{1}{2\pi \alpha^\prime} \int d^2\xi \, \sqrt{\gamma}\,,
\end{equation}
with $(\xi_1,\xi_2)$ the worldsheet coordinates and $\gamma$ the induced metric.
This method has been used to find the gluon condensate at $T=0$ in \cite{Andreev:2007vn}, while here this computation will be extended to small temperatures.
The first difference that emerges in this regime is that temporal and spatial Wilson loops are not equal, and correspond to different observables, namely the chromo-electric and chromo-magnetic components of the gluon condensate, respectively  \cite{Adami:1990sv}.

The model that will be used for this calculation is another version of the soft-wall model, in which the dilaton factor is inserted in the metric rather than in the action:
\begin{equation}
ds^2=\frac{e^{c_S^2 z^2}}{z^2} \left( f(z) d\tau^2 + d\bar x^2+\frac{dz^2}{f(z)}\right) \,\,\,;
\end{equation}
again, the scale mass $c_S$ can be set from the $\rho$ meson spectrum, obtaining $c_S=0.67$ GeV.

For small loops, we can expand the action \eqref{eq:NGaction} in terms of $\lambda=a^2 T^2$, with $a$ the radius of the circular loop (see Refs.~\cite{Colangelo:2013ila,Andreev:2007vn} for more details):
\begin{eqnarray}\label{eq:expansion}
 S_{NG}&=&S_0+\lambda S_1+\lambda^2 S_2 + {\cal O}(\lambda^3)\,,
\end{eqnarray}
finding $S_0=-1$, $S_1=5/3$, and $S_2^{\tau/y}=\frac{7}{90} \left(85\mp 2 \pi^4 \frac{T^4}{c_S^4} -120 \log 2\right)$.
$S_0$ and $S_1$ have the same expression for the spatial and temporal Wilson loop, while $S_2^\tau$ (from temporal WL) and $S_2^y$ (from spatial WL) have a different sign. By using Eqs.~\eqref{eq:logW},\eqref{eq:expNG},\eqref{eq:expansion} (we set $Z$=1), the electric ($e$) and magnetic ($m$) parts of the gluon condensate can be computed: 
\begin{equation}\label{eq:G2plaq}
G_2^{e/m}(T) = \frac{14 c_S^4}{5 \pi^4} \left(85\mp 2 \pi^4 \frac{T^4}{c_S^4} -120 \log 2\right) \,.
\end{equation}
This $T^4$ dependence is reliable only for small temperatures, since smaller and smaller values of the radius of the loop must be chosen to make the series \eqref{eq:expansion} convergent.
A similar behaviour has been found in Ref.~\cite{Eletsky1992rs}, while in Ref.~\cite{D'Elia2002ck} the magnetic component does not depend on $T$, and the electric one decreases with $T$.

\section{Concluding remarks}
Bottom/up holographic models can have a crucial impact in the field of strong interactions.
As an example, with little computational effort, finite temperature and also density effects can be studied, reproducing key features of QCD and getting important predictions.
In this work, a nonperturbative property of QCD, the gluon condensate, has been studied at finite temperature and density, a computation which is difficult to accomplish with other approaches. The model also predicts a first-order phase transition in thermodynamic functions at high values of chemical potential and small temperatures.

\section*{Acknowledgments}
I am grateful to P. Colangelo, S. Nicotri and F. Zuo for collaboration.

\end{document}